\documentclass{article}
\usepackage{spconf,amsmath,graphicx}

\usepackage{array}
\usepackage{multirow}
\usepackage{comment}
\usepackage{amsmath}
\usepackage{amsfonts}
\usepackage{graphicx}
\usepackage{xcolor}
\usepackage{color}
\usepackage{enumerate}
\usepackage{arydshln}
\usepackage{booktabs}
\usepackage{tikz}
\usepackage{algorithm}
\usepackage{subfig}
\usepackage{multirow}
\usepackage[noend]{algorithmic}
\usepackage{tablefootnote}
\usepackage{diagbox}
\usepackage{enumitem}

\title{Exploring Effective Distillation of Self-Supervised Speech Models for Automatic Speech Recognition}

\name{Yujin Wang$^{1,\dagger}$, Changli Tang$^{1,\dagger}$, Ziyang Ma$^2$, Zhisheng Zheng$^2$, Xie Chen$^{2,3, \ddagger}$, Wei-Qiang Zhang$^{1, \ddagger}$ \thanks{$^{\dagger}$ equal contribution, $^{\ddagger}$ corresponding authors}}
 \address{$^{1}$Department of Electronic Engineering, Tsinghua University\\ 
 $^2$ MoE Key Lab of Artificial Intelligence, AI Institute,  \\ X-LANCE Lab, Department of Computer Science and Engineering, \\ Shanghai Jiao Tong University, Shanghai, China \\
 $^3$ Peng Cheng Laboratory, Shenzhen, China
 }

\begin{document}
%
\maketitle

\begin{abstract}

Self-supervised learning (SSL) has achieved great success in speech processing, but always with a large model size to increase the modeling capacity. This may limit its potential applications due to the expensive computation and memory costs introduced by the oversize model. Compression for SSL models has become an important research direction of practical value. To this end, we explore the effective distillation of HuBERT-based SSL models for automatic speech recognition. First, a comprehensive study of different student model structures is conducted. On top of this, as a supplement to the regression loss widely adopted in previous works, a  discriminative loss is introduced for HuBERT to enhance the distillation performance, especially in low-resource scenarios. In addition, we design a simple and effective algorithm to distill the front-end input from waveform to Fbank feature, resulting in  17\%  parameter reduction and doubling inference speed, at marginal performance degradation.
\end{abstract}



\section{Introduction}
\label{sec:intro}

Recently, self-supervised learning (SSL) has made great progress in 
various areas including speech processing. The training of speech-based SSL is similar to that of text-based SSL, such as BERT \cite{devlin2018bert}, by reconstructing or predicting itself in an unsupervised way. Through effective SSL algorithms \cite{mohamed2022self}, the SSL model is able to encode rich hidden acoustic and semantic information implicitly from oceans of unlabeled data. A series of SSL algorithms have been developed and presented promising performance in a range of speech-related tasks~\cite{baevski2020wav2vec2,hsu2021hubert,chen2022wavlm,baevski2022data2vec,ma2022mt4ssl,chung2021w2v}. However, the large memory consumption and expensive computation load derived from the oversize SSL model might hinder their wide applications and deployments. 

Inspired by compression works~\cite{sanh2019distilbert,jiao2020tinybert} on BERT model in NLP domain, there are several previous studies on the distillation of SSL models in speech domain~\cite{chang2022distilhubert,wang2022lighthubert,lee2022fithubert,ashihara2022deep}, which attempts to reduce the model size for a well-trained SSL model in an unsupervised fashion. Most of these existing works are investigated on the SUPERB benchmark~\cite{yang2021superb}, a generic testing framework for pre-trained models on a range of downstream tasks. The SSL models are evaluated in the constrained track, where the 
whole upstream model is frozen, and the hidden layer outputs from the Transformer encoder are weighted for downstream tasks. The constrained configuration might not be able to fully reflect the potential of the distilled model for specific tasks. In addition, most SSL models adopt waveform as input, and several CNN layers are applied for feature extraction. This waveform-based front-end is retained after distillation in the student model, which is cumbersome in terms of efficiency and memory consumption. As a result, the existence of a waveform front-end might introduce potential issues for the deployment of this distilled model.

In this work, several aspects are investigated to facilitate the distillation of SSL models for speech recognition. The 
contributions of this paper can be summarized as follows:

    $\circ$ A comprehensive study of SSL model distillation is conducted to compare performances between different student model structures (e.g. deep \& thin, shallow \& wide) given the same amount of student model parameters, under unconstrained conditions.
    
    $\circ$ A discriminative loss is introduced to the distillation of HuBERT model in an unsupervised fashion, as a supplement and replacement of the layer-by-layer regression loss widely used in previous work. Significant performance improvements can be achieved by applying the proposed loss in low-resource scenarios.
    
    $\circ$ An effective distillation on front-end input is proposed and explored, which allows the student model to discard the troublesome waveform front-end and adopt a more practical Fbank front-end, with significant parameter reduction and inference speedup, at minor performance degradation.

\section{Related Works}
\label{sec:works}

\subsection{HuBERT}

HuBERT \cite{hsu2021hubert} is one of the most popular speech-based SSL models, attracting increasing research interest due to its simplicity and superior performance. During the training of HuBERT, the K-Means algorithm is applied to obtain the pseudo label for each frame, on top of the MFCC features or hidden layer outputs of a well-trained HuBERT model. The raw waveform is taken as input and fed into several CNN layers for feature extraction. After that, the feature is partially masked successively and encoded by Transformer layers, to predict K-Means pseudo labels $c$. The probability distribution for one masked frame in time $t$ can be computed as:

\begin{equation}\label{eq:prob}
    \mathbf{p}(c_t|X) = \frac{\exp(\cos(\mathbf{A}\mathbf{h}_t,\mathbf{e}_c)/\tau)}{\sum_{c^\prime=1}^C \exp(\cos(\mathbf{A}\mathbf{h}_t,\mathbf{e}_{c^\prime})/\tau)},
\end{equation}
where $\mathbf{e}_c$ is the embedding of the K-Means clustering label $c$, $\mathbf{h}_t$ represents the last hidden state of the Transformer 
encoder at time $t$, $\tau$ is the temperature to scale the logits, $\mathbf{A}$ is the projection matrix to generate the 
embedding of a hidden state, and $\cos (\cdot, \cdot)$ denotes the cosine similarity. The objective function of HuBERT shown below aims to minimize the cross entropy loss over pseudo labels of the masked frames:
\begin{equation}\label{eq:hubertloss}
    \mathcal{L} = -\sum_{t\in M} \log \mathbf{p} (c_t|X),
\end{equation}
where $M$ denotes the set of masked frames. 

\subsection{Distillation of SSL Models}
\label{sec:distill}

Knowledge distillation~\cite{gou2021knowledge} provides an effective solution to compress the model size \cite{hinton2015distilling}, which is also known as teacher-student learning \cite{jinyu2014teacherstudent} in literature. There are various ways for the student model to learn useful information from the teacher model. One typical approach is to minimize the Kullbac-Leibler divergence (KL-divergence) between the probability distributions of student and teacher models. Another alternative loss is to apply the L1 or L2 loss between the hidden layer outputs of student and teacher models \cite{adriana2015fitnets}. In most previous works, the knowledge distillation is operated in a supervised way like~\cite{yang2022knowledge}, where the ground truth label is given, and the student model learns from the teacher and ground truth simultaneously for better performance.  

In the realm of speech-based self-supervised learning, the distillation is normally conducted on unsupervised data. There are several successful attempts of applying knowledge distillation for speech based SSL models. The DistilHuBERT~\cite{chang2022distilhubert} model with shallow and wide (S\&W) structure is used to learn from the hidden layer outputs of the teacher model. \cite{wang2022lighthubert} proposes an efficient but complex two-stage distillation method. In \cite{lee2022fithubert}, the FitHuBERT model, with a deep and thin (D\&T) model structure, is able to achieve better results on content-related downstream tasks compared to~\cite{chang2022distilhubert}.
\cite{ashihara2022deep} presents a comprehensive 
study by comparing the performances between two different model architectures, S\&W and D\&T.
It is worth noting that both DistilHuBERT and FitHuBERT are investigated under the contrained track on the SUPERB benchmark, 
which might not be able to reflect the potential effect of the distilled model as it does not fully explore the powerful modeling capacity lied in the Transformer encoder and merely treats it as a frozen feature extractor during the whole fine-tuning stage, missing the opportunity to pursue strong but non-linear features~\cite{he2022masked}.

\section{Methods}
\label{sec:method}

Unlike most previous works~\cite{chang2022distilhubert,lee2022fithubert,ashihara2022deep}, where the SSL model distillation is investigated on a range of downstream tasks in the constrained track on the SUPERB benchmark, in this paper, we mainly focus on the ASR task, and the distillation model is fine-tuned under the unconstrained condition, i.e., the SSL model can also be updated to get the better performance. Given the superior performance of HuBERT and its popularity in previous distillation works, HuBERT is adopted for distillation in this work.

\subsection{Exploration of Student Model Structures}

\begin{figure}[htbp]
    \centering
    \vspace{-5pt}
    \includegraphics[width=\linewidth]{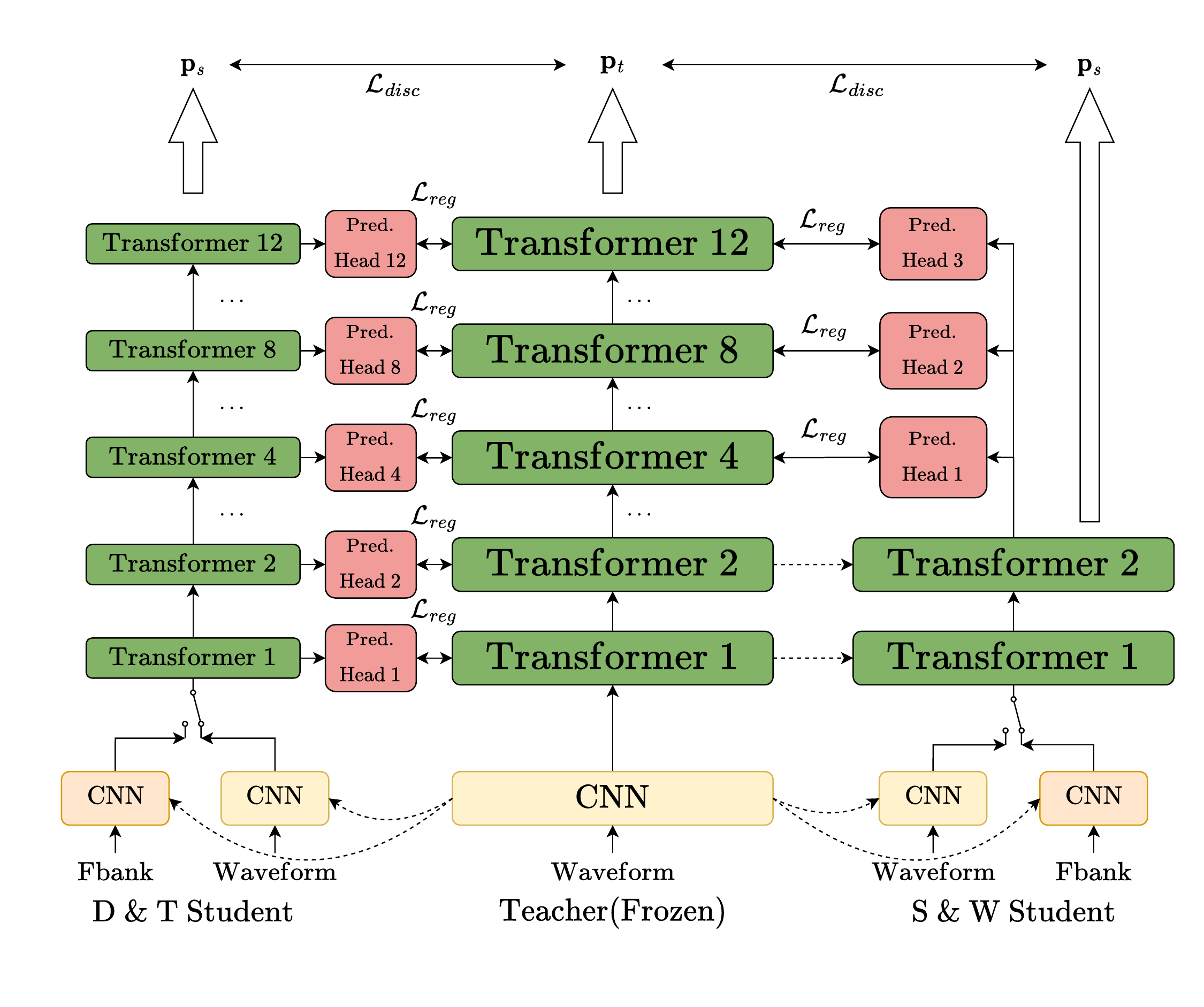}
    \vspace{-1.0cm}
    \caption{The distillation process of the two student structures is shown above, where the left side demonstrates the distillation of deep and thin (D\&T) model and the right side presents that of shallow and wide (S\&W) model with two optional front-ends (waveform or Fbank).\\}
    \label{fig:struct}
    \vspace{-0.8cm}
\end{figure}

Motivated by the investigation and findings in \cite{ashihara2022deep}, we first compare the performance of two different student model structures  with similar number of parameters under the unconstrained track, to determine the optimal model architecture for distillation. Figure \ref{fig:struct} illustrates the distillation process for these two student model structures.

\subsection{Discriminative Loss for SSL Model Distillation}
\label{subsec:distill}
As described in Section~\ref{sec:distill}, two types of information sources from the teacher model can be used for distillation, one is the probability distribution and the other is the hidden layer output. In the existing SSL model distillation algorithms, the latter information is widely adopted and the regression loss $\mathcal{L}_{reg}$ is applied to minimize the L1 or L2 distance of the hidden layer outputs between the student and teacher models. Inspired by the fact that pseudo labels are used as discrete targets in HuBERT and the cross entropy loss is applied for optimization,  it is straightforward to utilize the probability distribution calculated in the output layer of HuBERT for knowledge distillation. In order to make a distinction with the regression loss $\mathcal{L}_{reg}$ used in DistilHuBERT and FitHuBERT \cite{chang2022distilhubert,lee2022fithubert}, we name this distillation loss based on probability distribution as the discriminative loss $\mathcal{L}_{disc}$ .

These two types of distillation loss, $\mathcal{L}_{reg}$ and $\mathcal{L}_{disc}$, can be combined to derive the final distillation loss with different weights, which can be denoted as below:

\begin{equation}\label{eq:mainloss}
    \begin{split}
    \vspace{-0.2cm}
    & \mathcal{L}_{distill} = \lambda_{reg}\mathcal{L}_{reg} + \lambda_{disc}\mathcal{L}_{disc} \\
        & \mathcal{L}_{reg}  = \sum_{i\in \mathcal{T}}\{||\mathbf{h}_t^i - \mathbf{h}_s^i||_1 - \log\sigma[{\rm cos}(\mathbf{h}_t^i , \mathbf{h}_s^i)]\} \\
    & \mathcal{L}_{disc}  = {\rm KL}(\mathbf{p}_t, \mathbf{p}_s),
    \vspace{-0.2cm}
    \end{split}
\end{equation}
where $\lambda_{reg}$ and $\lambda_{disc}$ are set to tune the impact of regression and discriminative losses. 
For the distillation with the regression loss  $\mathcal{L}_{reg}$, we follow the previous works \cite{chang2022distilhubert,lee2022fithubert}. $\sigma$ denotes the sigmoid function, $\mathcal{T}$ denotes the set of teacher model's layers to learn. L1 and cosine similarity are adopted to close the gap between the hidden layer outputs of student and teacher models. For the discriminative loss $\mathcal{L}_{disc}$,  KL-divergence is applied to minimize the probability distribution distance between the student and teacher models. 
The probability distributions over the pseudo labels of teacher $\mathbf{p}_t$ and student $\mathbf{p}_s$ can be computed as follows:

\vspace{-0.5cm}
\begin{equation}\label{eq:mainprob}
    \begin{split}
        & \mathbf{p}_s = {\rm softmax}(\mathbf{W_p}\mathbf{h}_s^{last}) \\
        & \mathbf{p}_t = {\rm softmax}({\rm cos}(\mathbf{W_e}\mathbf{h}_t^{last}, \mathbf{l})/\tau),
    \end{split}
\end{equation}
where $\mathbf{h}_{t}^{last}$ and $\mathbf{h}_{s}^{last}$ are the last hidden layer outputs of the teacher and student models. The computation of $\mathbf{p}_t$ is the same as teacher HuBERT model, where $\tau$ is the temperature used in the teacher,  $\mathbf{W_e}$ is the embedding layer to produce the teacher's output embeddings, and $\mathbf{l}$ is the embeddings for K-Means pseudo labels. For the computation of $\mathbf{p}_s$, 
we adopt a linear output layer $\mathbf{W_p}$ followed by 
a softmax operation to compute the probability distribution for the student model.

\subsection{Distillation of Front-end Features}

\begin{figure}[htbp]
    \centering
    \vspace{-8pt}
    \includegraphics[width=\linewidth]{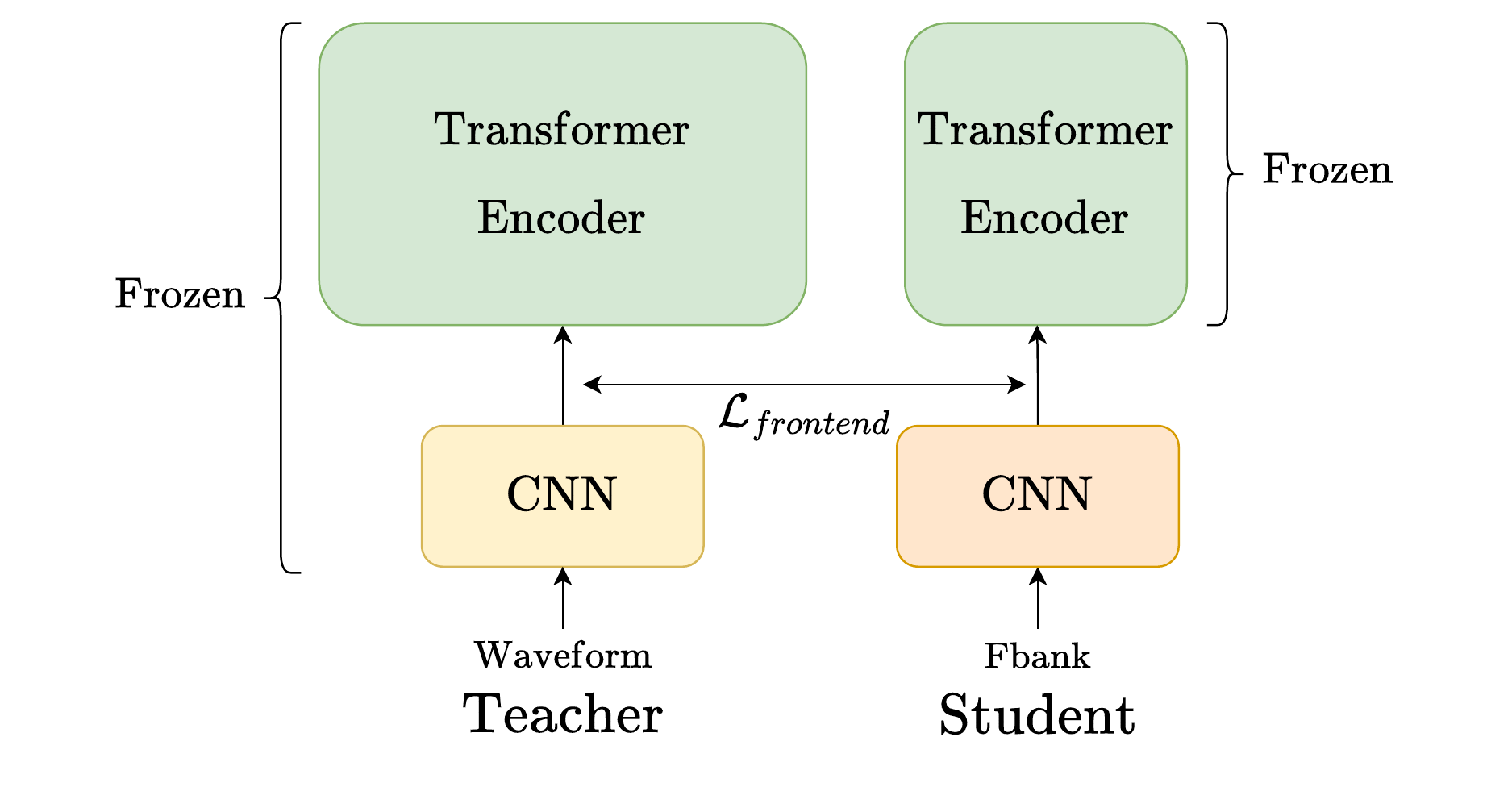}
    \vspace{-0.5cm}
    \caption{The front-end distillation during the first $N$ steps.}
    \label{fig:stage1}
    \vspace{-13pt}
\end{figure}

Due to the widespread influence of the Fairseq toolkit~\cite{ott2019fairseq} in the speech based SSL community, most publicly released SSL models follow the front-end configuration used in Fairseq, where the waveform is taken as input and a CNN module is applied for feature extraction. Conventional, previous distillation works on SSL models retain the waveform-based front-end in the student model, and mainly focus on the compression of the Transformer encoder. However, this waveform-based front-end takes a fair amount of parameters and is computationally expensive, especially when the backbone Transformer model shrinks largely in the student model. It turns out to be the potential bottleneck for the optimization of model size and efficiency. In order to address this issue, in addition to distilling the backbone Transformer model, the distillation of the front-end is explored. Instead of maintaining the waveform front-end, a more friendly Fbank front-end is adopted in the student model, which is distilled from the teacher model with the waveform front-end. We propose the following Fbank front-end distillation pipeline inspired by the front-end adapter adopted in the fine-tuning stage~\cite{chenadapter2022}:

\begin{equation}
    \mathcal{L} = 
\begin{cases}
	\mathcal{L}_{frontend}, & \text{if } {\rm step}\leq N\\
	\mathcal{L}_{distill},              & \text{otherwise},
\end{cases}
\end{equation}

where $N$ is the specified number of steps. For the first $N$ steps, only the front-end distillation is applied. Afterwards, the distillation is conducted as described in Section \ref{subsec:distill}.
$\mathcal{L}_{frontend}$ can be the L1 or L2 loss to minimize the distance between the outputs of the teacher's waveform front-end and the student's Fbank front-end. The pipeline of  $\mathcal{L}_{frontend}$ calculation is illustrated in Fig~\ref{fig:stage1}.

\section{Experiments}
\label{sec:exp}

\subsection{Experimental Setup}
\label{sec:expsetup}

\noindent \textbf{Model} \ The HuBERT Base model is used as the teacher, and two types of student model structures are investigated, which are shallow \& wide (S\&W) and deep \& thin (D\&T) respectively, as shown in  Table~\ref{tab:expparam}. 
For the student with waveform front-end, we keep the same front-end settings as the teacher model with seven CNN layers,  
while for the student with Fbank front-end, we select Fbank features with a frameshift of 10ms and downsample them to a frameshift of 20ms using only one CNN layer with a stride size of 2.

\noindent \textbf{Distillation} \ 
960-hour LibriSpeech dataset~\cite{panayotov2015librispeech} is used for knowledge distillation in an unsupervised fashion.
The default number of update is 25k, and other settings are kept the same as DistilHuBERT~\cite{chang2022distilhubert}. For the Fbank front-end student, we first apply the $\mathcal{L}_{frontend}$ loss for 5k steps, then the $\mathcal{L}_{distill}$ loss for 25k steps.
For the student with waveform front-end, the S\&W student is initialized with the teacher's front-end and the first two layers of the transformer encoder, while the D\&T student is only initialized with the teacher's front-end and the Transformer encoder is randomly initialized due to mismatch of dimension in Transformer between the teacher and student models. 
The S\&W student uses 3 prediction heads to respectively learn the teacher's 4th, 8th and 12th layers, while the D\&T student performs a layer-to-layer distillation over all layers.
 We change $\lambda_{disc}$ and $\lambda_{reg}$ in Equation \ref{eq:mainloss} to evaluate the distillation model with different settings.

\noindent \textbf{Fine-tuning} \ The fine-tuning experiments are implemented on Fairseq~\cite{ott2019fairseq} using CTC loss. 
The 1-hour \& 10-hour Libri-light \cite{librilight} and 100-hour LibriSpeech splits are used as fine-tuning data, with fine-tuning steps of 25k, 25k and 80k, and learning rates of 2e-5, 5e-4 and 5e-4 respectively.

\begin{table}[!htbp]
\centering
\small{
    \caption{Details of student model structures.}
    \begin{tabular}{c|c|c|c|c|c}
        \hline
        \multirow{2}{*}{\textbf{type}} & \multirow{2}{*}{\textbf{front-end}}  & \multirow{1}{*}{\textbf{\#param}} & \textbf{attn} & \textbf{FFN }   & \multirow{2}{*}{\textbf{\#layer }}   \\
        &  &(M)  &\textbf{dim} & \textbf{dim} &  \\
        \hline
         \multirow{2}{*}{S\&W}  & Waveform   & 23.64         & \multirow{2}{*}{768} & \multirow{2}{*}{3072} & \multirow{2}{*}{2}                                 \\
         & Fbank & 19.81         &                   &                    &                                          \\
        \hline
         \multirow{2}{*}{D\&T} & Waveform   & 23.08         & \multirow{2}{*}{480} & \multirow{2}{*}{480}  & \multirow{2}{*}{12}                           \\
          &Fbank & 19.25         &                   &                    &                                           \\
        \hline
    \end{tabular}%
    \vspace{-0.5cm}
    \label{tab:expparam}%
    }
\end{table}%

\subsection{Results}

The first experiment is to investigate the performance of different student model structures and distillation losses, and the results are shown in Table~\ref{tab:main_lo}. According to the results, we notice that the D\&T student model consistently outperforms the S\&W student model with similar amounts of parameters. 
This validates the previous findings \cite{lee2022fithubert,ashihara2022deep} reported on SUPERB under the constrained condition. Therefore, the D\&T model is more advantageous than the S\&W model for ASR and will be used as the baseline in the following experiments. It is also interesting to note that the improvement on the unconstrained condition (35.2\% relatively) is much larger than that on the constrained condition (9.6\% relatively) reported in \cite{ashihara2022deep} with 100-hour fine-tuning data and regression loss $\mathcal{L}_{reg}$. 

\begin{table}[ht]
    \centering
    \small{
    \caption{WER results of waveform front-end distilled models on LibriSpeech test-clean dataset with a 4-gram LM.}
    \begin{tabular}{ccccc}
        \toprule
        \textbf{structure}  & $\lambda_{reg},\lambda_{disc}$ & \textit{\textbf{1-hour}} & \textit{\textbf{10-hour}} & \textit{\textbf{100-hour}} \\

        \midrule

        S\&W  & 0,0                          & 98.73                    & 32.91                     & 14.19                      \\
        D\&T &         (w/o distillation)                   & 95.45                    & 64.49                       & 11.69                       \\
        \midrule
        S\&W  & \multirow{2}{*}{1,0}                           & 45.92                    & 24.64                     & 11.57                      \\
        D\&T &                           & 37.77                    & 17.82                     & \bf{7.50}                  \\
        \midrule
        S\&W  & \multirow{2}{*}{0,1}                           & 44.73                    & 21.98                     & \bf{11.23}                 \\
        D\&T &                            & 32.71                    & 17.17                     & 8.40                       \\
        \midrule
        S\&W  & \multirow{2}{*}{1,1}                           & \bf{39.42}               & \bf{21.86}                & 11.31                      \\
        D\&T &                            & \bf{30.67}               & \bf{16.83}                & 7.88                       \\

        \bottomrule 
    \end{tabular}
    \vspace{-6pt}
    \label{tab:main_lo}
    }
\end{table}

By introducing the discriminative loss, where $\lambda_{disc} = 1$, the ASR performance can be further improved compared to the baseline with regression loss, in most fine-tuning data and student model configurations, especially for the low-resource scenarios with 1-hour and 10-hour fine-tuning data. Even in the worse case, with the 100-hour fine-tuning data and the D\&T student model, the word error rate (WER) degradation caused by the additional discriminative loss is minor, which is from 7.50\% to 7.88\%.

The next experiment is to explore the distillation of front-end. Student models with the waveform front-end and Fbank front-end are compared. The D\&T student model is adopted for distillation because of its superior performance, and the statistics of student models can be found in Table \ref{tab:expparam}.
Table \ref{tab:fbank} gives the WER results of these two types of front-end. It can be seen that the student model with Fbank front-end can yield a decent WER performance after distillation, although it is slightly worse than the student model with waveform front-end. The degradation can be partially explained by the 17\% reduction of parameters in the student with Fbank front-end. It is also noteworthy that the introduction of discriminative loss consistently improves the ASR performance for student models with Fbank front-end in various amount of fine-tuning data. Table~\ref{tab:time} presents the elapsed inference time comparison for the student models with waveform and Fbank front-ends. A decomposition of consumed time on front-end and Transformer encoder is also given. According to the results, the inference speed of the student model with Fbank front-end is significantly faster than that of student with waveform front-end, in both 1 and 4 CPU threads, and a total inference speedup of 2 can be achieved by adopting the Fbank front-end in the student model.

\begin{table}[t]
  \centering
\small{
  \caption{WER of the D\&T student model with different front-ends on LibriSpeech test-clean set with a 4-gram LM.}
    \begin{tabular}{c|c|c|c|c}
    \hline
    \textbf{front-end} & \textbf{loss} & \textit{\textbf{1-hour}} & \textit{\textbf{10-hour}} & \textit{\textbf{100-hour}} \\
    \hline
    \multirow{2}[0]{*}{Waveform} & $\mathcal{L}_{reg}$ &  37.77   &  17.82    &  \textbf{7.50} \\
    &
           $\mathcal{L}_{reg} + \mathcal{L}_{disc}$ &  \textbf{30.67}   &  \textbf{16.83}    &  7.91  \\
    \hline
    \multirow{2}[0]{*}{Fbank} & $\mathcal{L}_{reg}$ &  32.08   &  19.42    &  8.49  \\ 
          & $\mathcal{L}_{reg} + \mathcal{L}_{disc}$ & \textbf{31.41} & \textbf{18.43} & \textbf{8.30}  \\
     \hline
    \end{tabular}%
    \vspace{-16pt}
  \label{tab:fbank}%
  }
\end{table}%


\begin{table}[htbp]
  \centering
  \small{
	\caption{Comparison of inference time between waveform and Fbank front-ends for the D\&T student model on CPU.}
    \begin{tabular}{c|cc|cc}
    \hline
    \multirow{3}[0]{*}{\textbf{stage}} & \multicolumn{4}{c}{\textbf{Inference time  (seconds)}} \\
         \cline{2-5} & \multicolumn{2}{c|}{\textbf{1 thread}} & \multicolumn{2}{c}{\textbf{4 threads}} \\
         \cline{2-5} & \textit{Waveform} & \textit{Fbank} & \textit{Waveform} & \textit{Fbank} \\
    \hline 
    front-end & 2055  & 32    & 684   & 14 \\
    Transformer & 2273  & 2130  & 834   & 781 \\
    \hline 
    \multirow{2}[0]{*}{total} & 4328  & 2162  & 1518  & 795 \\
    & (1.00X) & (2.00X) & (1.00X) & (1.91X) \\
    \hline
    \end{tabular}%
    \vspace{-9pt}
  \label{tab:time}%
  }
\end{table}%

\subsection{Details of Replacing Front-end}
In our approach, the student's Fbank features need to adapt to the teacher's waveform features first when distilling. Without this process, the performance of the student will be deteriorated a lot. Table~\ref{tab:fitfrontend} presents our results.
Adaptation to the front-end proves to be effective, probably because the features of the teacher's Transformers are adapted to the original waveform features. Using Fbank directly is more likely to result in a mismatch between the student's hidden states and the teacher's hidden states during distillation, reducing the distillation performance.

 Another discovery is that during fine-tuning, when $\mathcal{L}_{distill}$ and $\mathcal{L}_{frontend}$ are both used to update the Fbank front-end block, the gradient backpropagated from the $\mathcal{L}_{distill}$ loss may overwhelm the gradient of the $\mathcal{L}_{frontend}$ loss. So we divide the loss usage into two steps. We use Libri-light 10h as fine-tuning dataset and compare the experimental results in Table~\ref{tab:ablation}, which demonstrates the importance of loss settings.

\begin{table}[htbp]
	\centering
 \small{
	\caption{Comparison of whether to apply $\mathcal{L}_{frontend}$ first when distilling with $\mathcal{L}_{reg} + \mathcal{L}_{disc}$, evaluated on LibriSpeech test-clean dataset with a 4-gram LM.}
	\begin{tabular}{c|c|c|c}
		\hline
		{\diagbox{\textbf{method}}{\textbf{dataset}}} & \textit{\textbf{1-hour}} & \textit{\textbf{10-hour}} & \textit{\textbf{100-hour}} \\
		\hline
		w/o $\mathcal{L}_{frontend}$ &  34.90   &  19.08    &  9.21   \\
		w/ $\mathcal{L}_{frontend}$ &  \bf{31.41}   &  \bf{18.43}    &  \bf{8.30}   \\
		\hline
	\end{tabular}%
	\label{tab:fitfrontend}%
 }
\end{table}%

\vspace{-0.5cm}

\begin{table}[htbp]
\centering
\caption{Results of whether to divide the loss stage, evaluated on LibriSpeech test-clean dataset.}
\begin{tabular}{c|c}
    \hline
    \textbf{method} & \textbf{WER} \\
     \hline
     30k steps $\mathcal{L}_{frontend} + \mathcal{L}_{distill}$  & 23.09 \\
     5k steps $\mathcal{L}_{frontend}$, 25k steps $\mathcal{L}_{distill}$ & 18.43 \\
     \hline
\end{tabular}\label{tab:ablation}
\end{table}

\vspace{-0.2cm}

\subsection{About the Teacher Model}
Theoretically, a teacher model can also be a fine-tuned SSL model. We distilled HuBERT which tuned on LibriSpeech 960h into S\&W structure, and fine-tuned it on Libri-light 10h dataset. We list the following WER results in Table \ref{tab:distill_ft_model}: distilling a pretrained model then fine-tuning (DT+FT), distilling a fine-tuned model (FT+DT), and another fine-tuning for the distilled model on Libri-light 10h (FT+DT+FT).

Our experiments in Table \ref{tab:distill_ft_model} show that distilling a fine-tuned model is not directly usable, and further fine-tuning on the relevant data is required. This is more resource intensive while the final result is similar to directly distilling a pre-trained model.

\begin{table}[htbp]
  \centering
  \small{
      \caption{Effect of the order of fine-tuning and distilling}
    \begin{tabular}{c|c|c|c}
        \hline
        \textbf{Distilled S\&W Model} & DT + FT & FT + DT & FT + DT + FT \\
        \hline 
        \textbf{WER} & 11.23 & 25.89 & 11.02 \\
    \hline
    \end{tabular}%
    \vspace{-0.7cm}
  \label{tab:distill_ft_model}%
  }
\end{table}%

\subsection{Feature Analysis}

\begin{figure}[ht]
	\centering
	\caption{CCA similarity of hidden states between D\&T models before and after fine-tuning on low-resource Libri-light fine-tuning datasets (1h \& 10h).
     We choose 100 audio clips in LibriSpeech test-clean dataset for analysis.
     }
	\vspace{-10pt}
    \includegraphics[width=\linewidth]{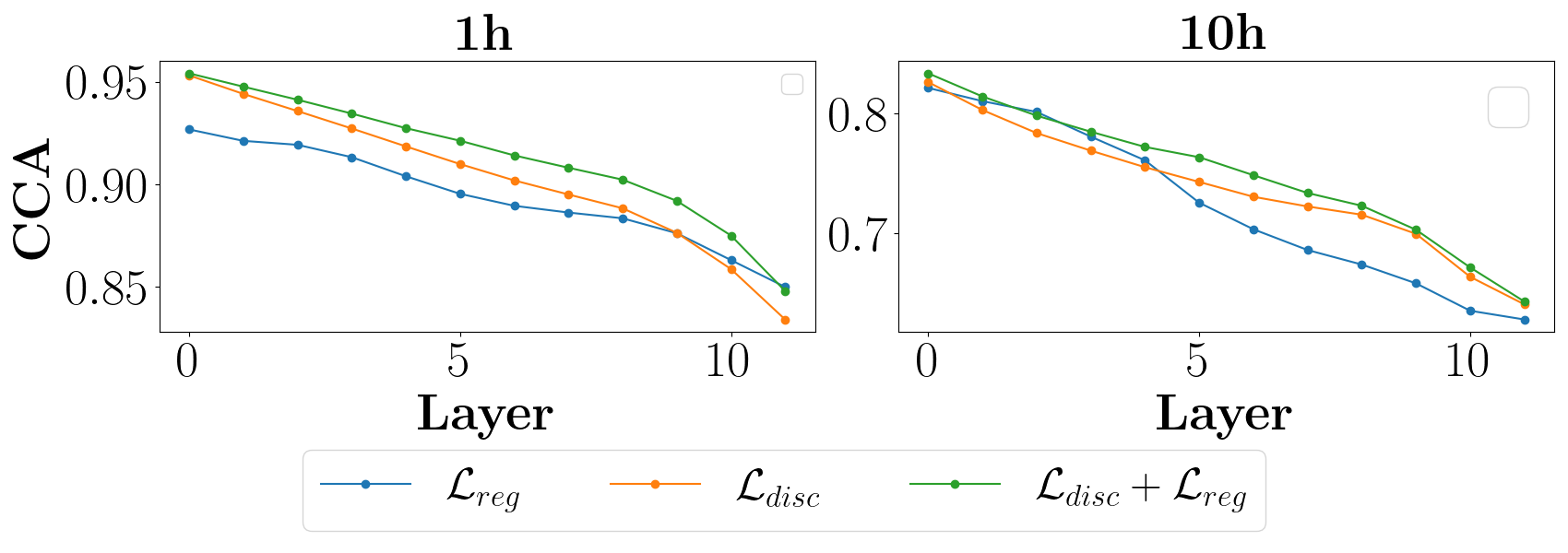}
	\label{fig:cca}
    \vspace{-0.9cm}
\end{figure}

Zhu et al.~\cite{zhu2022wav2vec} identified the ASR capability of a pre-trained model by analysing the CCA similarity~\cite{morcos2018insights,pasad2021layer}, reasoning that the higher the CCA similarity, the less the model changed during fine-tuning, indicating a more capable model.
Based on this, we calculate the CCA similarity for the distillation models before and after fine-tuning.
As Figure~\ref{fig:cca} shows, models distilled with $\mathcal{L}_{disc}$ have more similar representation to the fine-tuned model at low-resource cases,
which implements that $\mathcal{L}_{disc}$ is able to build better linguistic representational ability under low-resource scenarios for distilled models.

\section{Conclusion}
\label{sec:conclusion}

In this work, a series of explorations are conducted on the SSL model distillation for ASR.
We validate that the deep and thin structure is more advantageous to ASR given limited model parameters.
As a supplement to the regression loss used in previous work, 
a discriminative loss is introduced into the distillation of HuBERT in an unsupervised fashion, which significantly improves the performance on low-resource ASR.
Furthermore, we propose a front-end distillation framework by replacing the waveform with the Fbank feature as the input of the student model. This can reduce 17\% of the parameters and halve the inference time at the cost of minor performance degradation. 

\vspace{-0.2cm}
\section{Acknowledgement}
\label{sec:acknowledgement}
This work was supported by the National Natural Science Foundation of China under Grant No.~62206171 and No.~62276153, and the International Cooperation Project of PCL, and Alibaba Group through Alibaba Innovative Research Program.

\bibliographystyle{IEEEtran}
\bibliography{strings,refs}

\end{document}